\documentclass[twocolumn,epsf,showpacs,prd]{revtex4}
\usepackage{graphicx}
\usepackage{dcolumn}
\usepackage{bm}
\begin{document}

\newcommand{\be}{\begin{equation}}
\newcommand{\ee}{\end{equation}}
\newcommand{\ba}{\begin{eqnarray}}
\newcommand{\ea}{\end{eqnarray}}
\def\bone{B^{(1)}}

\title{Indirect Detection of Kaluza--Klein Dark Matter}
\author{Gianfranco Bertone$^a$, G\'eraldine Servant$^{b,c}$,
G{\"u}nter Sigl$^a$}
\affiliation{$^a$ GReCO, Institut d'Astrophysique de Paris, C.N.R.S.,
98 bis boulevard Arago, F-75014 Paris, France\\
$^b$ Enrico Fermi Institute, University of Chicago, Chicago, 
IL 60637\\
$^c$ High Energy Physics Division, Argonne National 
Laboratory, Argonne, IL 60439}

\vspace{0.5truecm}
\begin{abstract}
We investigate prospects for indirect 
detection of Kaluza--Klein dark matter, focusing on the
annihilation radiation of the first Kaluza--Klein excitation 
of the Hypercharge gauge boson $B^{(1)}$
in the Galactic halo, in particular we estimate neutrino, 
gamma-ray and synchrotron fluxes. Comparing the predicted
fluxes with observational data we are able to constrain
the $B^{(1)}$ mass (and therefore the compactification scale). The 
constraints depend on the
specific model adopted for the dark matter density profile.
For a NFW profile the analysis of synchrotron radiation 
puts a lower bound on the $B^{(1)}$ mass of the order of 
$\simeq300\,$GeV.
\end{abstract}

\pacs{04.50.+h, 12.60.-i, 14.80.-j, 95.35.+d, 96.40.-z \hfill ANL-HEP-PR-02-099,
EFI-02-52}
\maketitle

\vspace{1truecm}

\section{Introduction}

Over the last thirty years, independent pieces of evidence have 
accumulated in favor of the existence of \textit{Dark Matter} (DM), 
indicating that most of the matter of the universe is non baryonic and of
unknown nature. Particle physicists have come up with various 
DM candidates. The most promising and most extensively studied is the 
so-called \textit{neutralino}, the Lightest Supersymmetric Particle (LSP)
 which arises in supersymmetric models and is stable in models with
conserved R-parity. 

Although theoretically well motivated, 
\textit{neutralinos} are not the only viable DM candidates and it 
is important to investigate other possibilities arising in extensions 
of the Standard Model which are not the Minimal Supersymmetric Standard
Model (MSSM). For instance, models with compact extra dimensions possess
plenty of new states, Kaluza--Klein (KK) particles. Very
different types of models with extra dimensions exist, each showing a 
distinctive phenomenology. In \textit{brane world} scenarios, the 
Lightest Kaluza-Klein Particle (LKP) is not stable and cannot be a DM 
candidate. However, in models with \textit{Universal Extra Dimensions} (UED)
\cite{Appelquist:2000nn}, in 
which all Standard Model fields propagate in extra dimensions, including 
fermions, the LKP may be stable if 
\textit{KK parity} is preserved. \textit{KK parity} is preserved 
in all models where boundary lagrangians are identical at both orbifold 
fixed points. Therefore, in this general class of models, the LKP
turns out to be stable and was recently shown to be a viable DM 
candidate \cite{Servant:2002aq}. 
 
The LKP is likely to be associated with the first KK excitation of the
photon, more precisely the first KK excitation of the Hypercharge gauge boson 
\cite{Cheng:2002iz}, and we will refer to it as $\bone$. The relic
density of $\bone$ was computed 
in Ref.~\cite{Servant:2002aq} where 
it was shown that if the LKP is to account for DM then its mass (which is 
inversely proportional to the compactification radius $R$) should lie in 
the range 400--1200 GeV which is above any current experimental constraint: In
the case of one extra dimension, the 
constraint on the compactification scale in UED models from precision
electroweak measurements is as low as $R^{-1}\gtrsim 300$ GeV 
\cite{Appelquist:2000nn}. Very recently it was even argued that this
bound can be weakened to $R^{-1}\gtrsim 280$ GeV if one allows a Higgs mass 
as heavy as $m_H\gtrsim 800$ GeV \cite{Appelquist:2002wb}.
This is to be contrasted with another class of models 
where SM bosons propagate in extra dimensions while fermions are localized
in 4 dimensions, in such cases, the constraint on
the compactification scale is much stronger ($R^{-1}\gtrsim$ several 
TeV \cite{Cheung:2001mq}).

Direct detection of the LKP via its elastic scattering with nuclei in
a detector was investigated in Ref.~\cite{Cheng:2002ej,Servant:2002hb}.
It was emphasized in \cite{Servant:2002hb} that one-ton detectors are 
needed to probe the expected heavy masses as indicated by the
relic density calculation \cite{Servant:2002aq} of the LKP and one has
to wait for the next generation of direct detection experiments such
as GENIUS \cite{Klapdor-Kleingrothaus:2000eq}
or XENON \cite{Aprile:2002ef}. Simultaneously, LHC 
should probe most of the relevant KK mass parameter space (up to
$R^{-1}\sim 1.5$ TeV \cite{Cheng:2002ab})
and definitively confirm or rule out UED at the TeV scale.

In addition, there are other ways to probe KK DM \textit{indirectly}: As for 
other Weakly Interacting Massive Particles (WIMPs), the LKP could be
detected via the measurement of fluxes of particles coming from its
self annihilation. WIMPs annihilate with higher probability in regions
where the density of DM is higher, like in the Sun or in the center of
the Galaxy. Indirect detection prospects from LKP annihilation in the
Sun was investigated in \cite{Cheng:2002ej,Hooper:2002gs}. The idea 
is to look for a neutrino spectrum which could be detected by IceCube. 
Annihilation in the Galaxy can give rise to different fluxes:
Positron excess from LKP annihilation was investigated in
Ref.~\cite{Cheng:2002ej}. While the signal (which could be detected
by AMS) is spectacular for KK masses below 500 GeV, it is almost
indistinguishable from background for masses above 800 GeV. The photon
spectrum was investigated as well in Ref.~\cite{Cheng:2002ej} and
is also studied in the present article.

The aim of this paper is to investigate the prospects for indirect 
detection of KK dark matter, focusing on radiation from the
annihilation of such particles in the Galactic halo. In
particular we estimate neutrino and gamma-ray fluxes, 
and the synchrotron radiation from $e^+e^-$ pairs propagating
in the Galactic magnetic field. The paper is organized as 
follows: In section \ref{prof} 
the density profiles of dark matter in galaxies and
in particular in the Milky Way are presented. Section \ref{cross} 
discusses the annihilation of $B^{(1)}$ particles and the
details of fragmentation of secondary quarks. In section 
\ref{gamma} we evaluate gamma and neutrino fluxes from the
galactic center and compare the predicted fluxes with observations.
In section \ref{sy} we estimate the synchrotron radiation 
from $e^+e^-$ propagating in the Galactic magnetic field 
and the constraints on the $B^{(1)}$ mass. Finally, 
results are summarized and discussed
in section \ref{conc}.

\section{Dark Matter Profiles}
\label{prof}

In this section we discuss some widely used profile models for
the density of dark matter in galaxies, and
the corresponding value for the integration along the line 
of sight in the direction of the Galactic Center (GC).

The choice of the dark matter density profile is crucial 
when discussing annihilation radiation, because it fixes
the normalization of the observed spectrum. Even under the 
simplifying assumption of a spherically symmetric profile,
the uncertainty on the dark matter distribution is such 
that it is impossible to put model-independent constraints
on physical parameters of dark matter particles. 

In fact, there is still no consensus about the shape of 
dark halos. High-resolution N-body simulations suggest the
existence of ``cusps'', with the inner part of the halo density
following  a power law $\propto r^{-\gamma}$ with index $\gamma$
possibly as high as 1.5 (see below).
On the other hand observations of rotation curves of galaxies
seem to suggest much shallower inner profiles \cite{salucci} 
(but other groups claim the impossibility of constraining dark 
matter with such observations \cite{vandenBosch:2000rz}).

For what concerns the Milky Way, the situation is unclear despite 
the wide range of observational data available. Binney \& Evans 
(BE, 2001)~\cite{Binney:2001wu}
exclude cuspy profiles with $\gamma>0.3$, with an analysis based on
micro-lensing optical depth. Nevertheless Klypin, Zhao \& Somerville
(KZS, 2001)~\cite{Klypin:2001xu} find a good agreement between
Navarro, Frenk, and White (NFW) profiles, $\gamma=1$, and observational
data for the Galaxy and M31. The main difference between the 
two analysis is in the modelisation of the Galaxy: KZS claim
to have taken into account dynamical effects neglected by BE, 
and to have a ``more realistic'' description of the bar.     

The usual parametrisation for the dark matter halo density is
\begin{equation}
  \rho(r)= \frac{\rho_0}{(r/R)^{\gamma}
  [1+(r/R)^{\alpha}]^{(\beta-\gamma)/\alpha}}\label{profile}
\end{equation}
In Tab.~\ref{tab} we give the values of the respective parameters for
some of the most widely used profile models, namely the Kravtsov et al.
(Kra, \cite{kra}), 
Navarro, Frenk and White (NFW, \cite{Navarro:1995iw}), Moore et al. 
(Moore, \cite{Moore:1999gc}) and modified isothermal (Iso, e.g. 
\cite{Bergstrom:1997fj}) profiles.

The dark matter profile in the inner region of the Milky Way is 
even more uncertain: observations of velocity dispersion of high 
proper motion stars suggest the existence of a Super 
Massive Black Hole (SMBH) lying at the centre of our Galaxy, with a mass 
$\approx 2.6 \times 10^6 M_\odot$ ~\cite{Ghez:1998ab}. 

It has been argued ~\cite{Gondolo:1999ef} that the process of adiabatic 
accretion of dark matter on the central SMBH would produce a ``spike''
in the dark matter density profile, leading to a power law index possibly as
high as $\gamma \approx 2.4$. Although central spikes could be destroyed
by astrophysical processes such as hierarchical mergers 
~\cite{Ullio:2001fb, Merritt:2002vj}, these dynamical destruction 
processes are unlikely to have occurred for the Milky Way~\cite{Bertone:2002je}.
The existence of such spikes would produce a dramatic enhancement of the 
annihilation radiation from the GC, and would allow to put stringent
constraints on dark matter particles properties and distribution
~\cite{Gondolo:1999ef, Gondolo:2000pn, Bertone:2001jv, Bertone:2002je}.  

As a first step for the study of indirect detection of KK dark matter, we 
choose to be conservative, and to focus on ordinary profiles without central 
spikes.


We now want to compute the observed flux from dark matter
particle annihilation in the GC. The observed
flux can be written as
\begin{equation}
\Phi_i(\psi,E)=\sigma v \frac{dN_i}{dE} \frac{1}{4 \pi M^2}
\int_{\mbox{line of sight}}d\,s
\rho^2\left(r(s,\psi)\right)\label{flux}
\end{equation}
where the index $i$ denotes the secondary particle observed 
(we focus on  $\gamma-$rays and neutrinos) and the
coordinate $s$ runs along the line of sight, in a
direction making an angle $\psi$ respect to the direction
of the GC. $\sigma v$ is the annihilation cross 
section, $dN_i/dE$ is the spectrum of secondary particles
per annihilation and $M$ is the mass of the LKP.

\begin{table}
\caption{\label{tab} Parameters of some widely used profile models for 
the dark matter density in galaxies in Eq.~(\ref{profile}). We also
give the value of $\overline{J}(10^{-3})$, see text for details.}
\begin{ruledtabular}
\begin{tabular}{cccccc}
&$\alpha$&$\beta$&$\gamma$&R (kpc)&$\overline{J}\left( 10^{-3}\right)$ \\
\hline \\
Kra& 2.0& 3.0&0.4 & 10.0 &$ 2.166 \times 10^1$ \\
NFW& 1.0& 3.0& 1.0& 20& $1.352 \times 10^3$\\
Moore& 1.5& 3.0& 1.5& 28.0 &$ 1.544 \times 10^5$ \\
Iso& 2.0& 2.0& 0& 3.5&$2.868 \times 10^1$\\
\end{tabular}
\end{ruledtabular}
\end{table}

In order to separate the factors depending on the profile 
from those depending only on particle physics, we introduce,
following \cite{Bergstrom:1997fj}, the quantity $J(\psi)$
\begin{equation}
J\left(\psi\right) = \frac{1} {8.5\, \rm{kpc}} 
\left(\frac{1}{0.3\, \mbox{\small{GeV/cm}}^3}\right)^2
\int_{\mbox{\small{line of sight}}}d\,s\rho^2\left(r(s,\psi)\right)\,.
\end{equation}
We define $\overline{J}(\Delta\Omega)$ as  the average of $J(\psi)$ over
a spherical region of solid angle $\Delta\Omega$, centered on $\psi=0$.
The values of $\overline{J}(\Delta\Omega=10^{-3})$ are shown in the
last column of table \ref{tab} for the respective density profiles. 

We can then express the flux from a solid angle $\Delta\Omega$ as
\ba
\nonumber
\Phi_{i}(\Delta\Omega, E)\simeq5.6\times10^{-12}\frac{dN_i}{dE} 
&\left( \frac{\sigma v}
{\rm{pb}}\right)\left( \frac{1\rm{TeV}} 
{\rm{M}}\right)^2 \overline{J}\left(\Delta\Omega\right) \\
& \times \; \Delta\Omega\,\rm{cm}^{-2} \rm{s}^{-1}\,.
\label{final}
\ea

\section{Cross Sections and Fragmentation Functions}
\label{cross}

Annihilation cross sections of the LKP into Standard Model particles 
can be found in the appendix of \cite{Servant:2002aq}. Here we are concerned 
with the annihilation into fermions $f$ which, in the non relativistic 
expansion limit ($\langle \sigma v \rangle \simeq a+bv^2$), is given by:
\be
\sigma v(\bone \bone\rightarrow f\overline{f})
=\frac{{\alpha_1}^2\,N_c\, {N_f}\pi \,Y^4}{9\,M^2} (8-v^2)  
\ee
$N_c$, $N_f$ and $Y$ are respectively the number of colors, number of generations and 
hypercharge of
fermion $f$. For neutrinos, we obtain
\begin{equation}
(\sigma v)_{B^{(1)}B^{(1)}\rightarrow \nu\overline{\nu}} =
1.74313 \times 10^{-5}\:\: M^{-2}
\end{equation}
per neutrino flavor.
In addition to their direct production, neutrinos can also be 
produced via subsequent decay of charged pions $B^{(1)}B^{(1)}
\rightarrow q\overline{q} \rightarrow \pi^{\pm}\rightarrow \nu+..$. 
Direct annihilation into two $\gamma-$rays is loop-suppressed, so 
high energy photons are mainly produced from decaying
neutral pions $B^{(1)}B^{(1)}\rightarrow q\overline{q} \rightarrow \pi^0
\rightarrow \gamma \gamma$.
To investigate the spectrum of secondary neutrinos and 
$\gamma-$rays, one has to go through the details of quark 
fragmentation into (neutral or charged) pions and their 
subsequent decay.
\begin{figure}
\includegraphics[width=0.48\textwidth,clip=true]{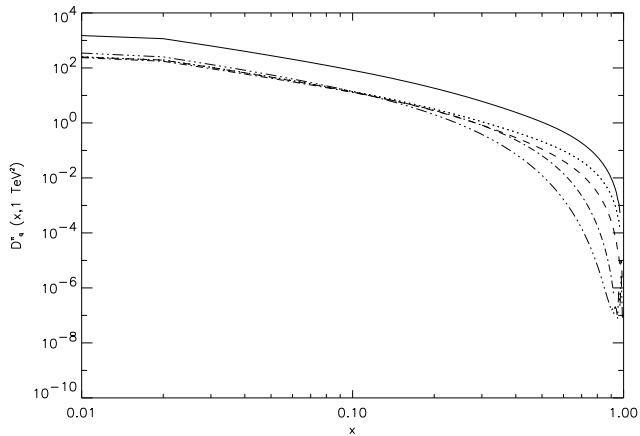}
\caption{Fragmentation function of different quark species into pions.
 From bottom to top, the curves are relative to quarks c and b (on the same 
curve, three dots-dashed line), s (dash-dotted), d (dashed), u (dotted) 
and sum of all (solid).}
\label{fragmall}
\end{figure}

We use the Fragmentation Functions (FFs) described
by Kretzer (2000) and implemented in \cite{Kretzer:2000yf}.
In this approach the hadronisation of partons is described
by the function $D^h_a(x,Q^2)$, which is the probability 
that the parton $a$ fragments into a hadron $h$ carrying
a fraction $x$ of the total momentum $Q$, where we sum over
partons and anti-partons. For the processes we are considering
$Q^2\simeq M^2$.

In Fig.~\ref{fragmall} we show the FFs relative to different quarks for the
production of Pions. 

\begin{figure}
\includegraphics[width=0.48\textwidth,clip=true]{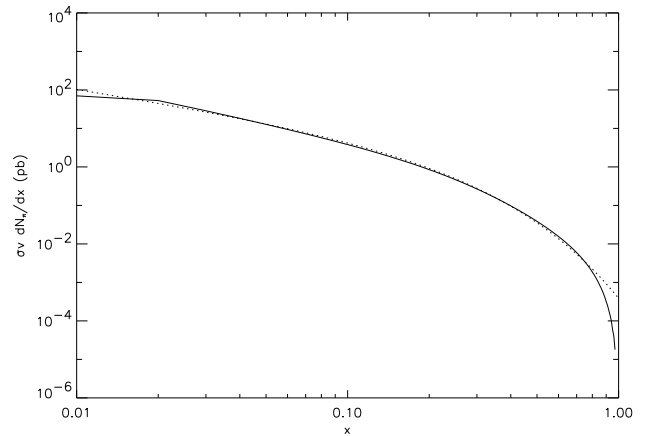}
\caption{Spectrum of charged pions after fragmentation times cross
section (see Eq.~(\ref{toth}), with  $h=\pi^\pm$ and $Q^2=1\,{\rm TeV}^2$). 
The dotted line is the analytic fit Eq.~(\ref{fit})which is sufficiently 
accurate up to $x\simeq 0.8$.}
\label{fragfit}
\end{figure}

The spectrum of hadrons $h$ produced after fragmentation is
the sum of the FFs of the different quarks, weighted
with the corresponding quark pair production cross sections in 
$B^{(1)}$ self-annihilation,
\begin{equation}
\sigma v\frac{dN_h}{dx}= \sum_a \sigma_av\;D^{h}_a (x,Q^2)\,.\label{toth}
\end{equation} 
The result for charged pions, $h=\pi^\pm$, is shown in Fig.~\ref{fragfit}
for $Q^2=1\,{\rm TeV}^2$, along with the analytic fit
\begin{equation}
f(x) \simeq \frac{0.7}{x^{1.1} e^{7.5\;x}}\,\mbox{pb}\,,\label{fit}
\end{equation}
which is sufficiently accurate up to $x \simeq 0.8$.
 
We are now able to compute the $\gamma-$ray spectrum resulting 
from neutral pion decay into two photons. The spectrum of 
photons from the decay of a single neutral pion is flat,
\begin{equation}
\frac{dN}{dE_\gamma}=\frac{2}{P_{\pi}}   
\end{equation}
for $E_\gamma$ in the interval $(E_{\pi} \pm P_{\pi})/2$, where
$E_i$ and $P_i$ denote energy and momentum of the respective
particle.

For relativistic pions, $E_{\pi} \simeq P_{\pi}$, so that 
the spectrum can be approximated with a Heavyside function.
The $\gamma-$ray spectrum is thus 
\begin{equation}
\frac{dN_{\gamma}}{dx}=\int_{0}^{1}f(x^\prime)
\frac{1}{x^\prime}\theta(x^\prime-x)dx^\prime\,.
\end{equation}
Examples of predicted fluxes for different particle masses
and different density profiles are given in the next section.

\begin{figure}
\includegraphics[width=0.48\textwidth,clip=true]{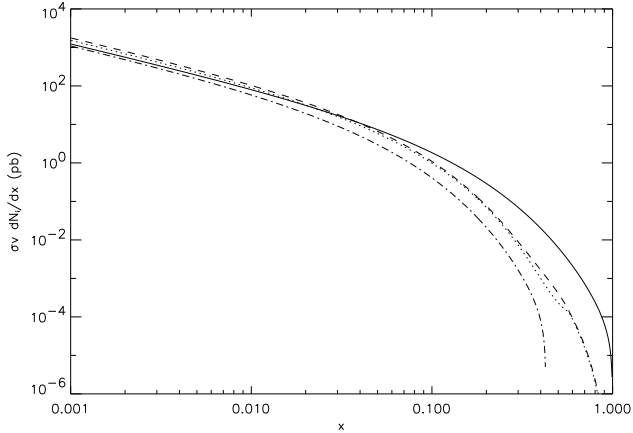}
\caption{The spectra of $\gamma-$rays (solid), $e^+$ (dashed),
$\overline\nu_{\mu}$ (dotted), $\nu_e$ and $\nu_{\mu}$ (dash-dotted), 
resulting from folding Eq.~(\ref{toth}) with the pion decay
spectra.}
\label{gaenu}
\end{figure}

Finally, to compute the neutrino spectra we use the formulae 
provided in~\cite{Lee:1996fp}. In Fig.~\ref{gaenu} we show the
resulting spectra of $\gamma-$rays, $e^+$, $\overline{\nu}_\mu$,
${\nu}_\mu$, and ${\nu}_e$.

\section{Gamma-Ray and Neutrino Fluxes}
\label{gamma}

\begin{figure}
\includegraphics[width=0.48\textwidth,clip=true]{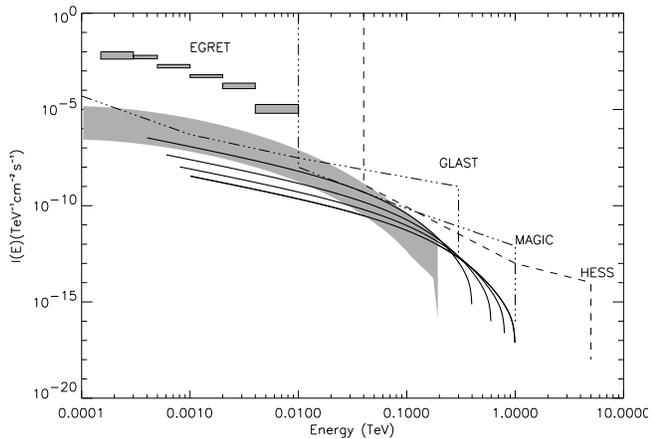}
\caption{Expected $\gamma-$ray fluxes for (top to bottom) $M=0.4$, 0.6, 0.8,
and 1 TeV and $\overline{J}\left(10^{-3}\right) = 500$. For
comparison shown are typical $\gamma-$ray fluxes predicted for
neutralinos of mass $\simeq200\,$GeV, as well as EGRET data and expected 
sensitivities of the future GLAST, MAGIC and HESS experiments.}
\label{gammas}
\end{figure}

In Fig.~\ref{gammas} we show the expected $\gamma-$ray flux
in a solid angle $\Delta\Omega=10^{-3}$ in the direction
of the GC for $M=0.4$, 0.6, 0.8, and 1 TeV  and for
$\overline{J} \left( 10^{-3} \right) = 500$. To obtain the flux
for a given profile, one can use the corresponding value of 
$\overline{J} \left( \Delta\Omega \right)$ given in the last column
of Tab. ~\ref{tab}. In the same figure we show for comparison
observational data from 
EGRET \cite{mayer}, and expected sensitivities of the future experiments 
GLAST \cite{Sadrozinski:wu}, MAGIC \cite{Petry:1999fm} and 
HESS~\cite{Volk:2002iz}.

\begin{figure}
\includegraphics[width=0.48\textwidth,clip=true]{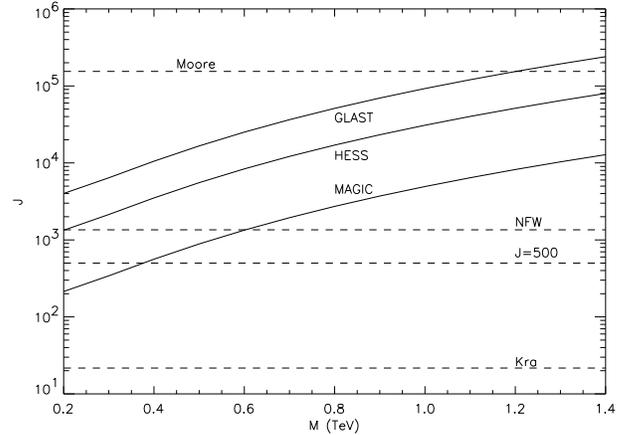}
\caption{Value of $J=\overline{J}(10^{-3})$ required to produce $\gamma$ fluxes
observable by the future GLAST, MAGIC and HESS experiments, as a function of
the $\bone$ mass. For comparison we show the value of $J$ for some 
profiles discussed in the text.}
\label{constr}
\end{figure}

We notice that Eq.~(\ref{final}) can be written as
\ba
\nonumber
\Phi_{l}(\Delta\Omega)  \simeq  5.6 \times 10^{-12}&\frac{dN_l}{dE} 
\left( \frac{a}
{\rm{pb}}\right)\left( \frac{1\rm{TeV}} 
{M}\right)^4  \overline{J}\left(\Delta\Omega\right)\\
& \times \; \Delta\Omega \ \rm{cm}^{-2} \ \rm{s}^{-1}\,.
\label{fluxM}
\ea
In this formula we used the fact that $B^{(1)}$ particles are
expected to be non-relativistic in the GC, so
we can safely use the non-relativistic limit of the cross section  
$\sigma v \rightarrow a (M / \mbox{\small{1TeV}})^{-2}$.

\begin{figure}
\includegraphics[width=0.48\textwidth,clip=true]{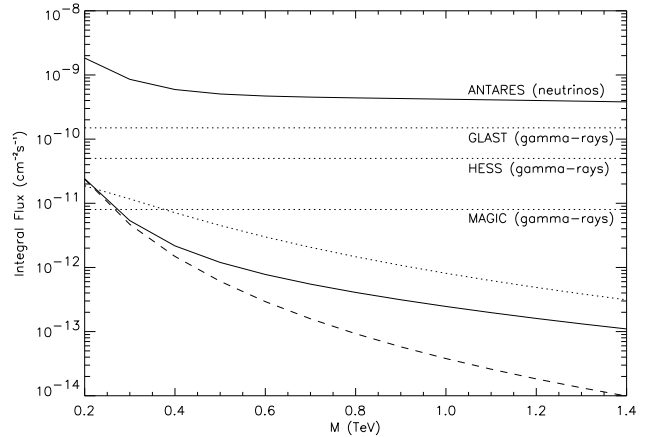}
\caption{Integral flux of $\gamma-$rays (dotted line) and 
muon neutrinos (solid) above 50 GeV, for 
$\overline{J}\left(10^{-3}\right) = 500$ . The dashed line shows 
the contribution of direct $\bone$ annihilation into neutrinos.
Horizontal lines are sensitivities 
of present and future experiments for $\gamma-$rays (dotted horizontal
lines) and neutrinos (upper solid line).}
\label{intflux}
\end{figure}

Given the particle physics details (cross sections and fragmentation)
we are thus left with two free parameters: the mass of the dark
matter particle, $M$, and the value of $\overline{J}(\Delta\Omega)$, depending
on the specific dark matter profile adopted. We show in
Fig.~\ref{constr} the constraints on these two parameters
based on the expected sensitivity of GLAST, MAGIC and HESS. 
For a NFW profile masses below 600 GeV are excluded if MAGIC does 
not observe any radiation from the GC.

Constraints from neutrino fluxes are weaker. High energy
neutrinos can be detected with large underground telescopes,
which are sensitive to muons originated by 
charged current interactions of $\nu_\mu$ with the matter
surrounding the detector. 
 
In Fig.~\ref{intflux} we plot the integral flux of muon neutrinos 
above $50$ GeV (solid line) as a function of the $\bone$ mass.
This flux is obtained by adding the contribution of neutrinos
from three different channels:
\begin{itemize}
\item neutrinos produced directly by $\bone$ annihilations 
(dashed line), their spectrum being a line at energy $E=M$ 
\item secondary neutrinos from decay of charged pions. This
spectrum can be evaluated using the above cited expressions
for the charged pion decay obtained in Ref.~\cite{Lee:1996fp}
\item secondary neutrinos from ``prompt'' semi-leptonic
decay of heavy quarks. This spectrum is given, for example, in
Ref.~\cite{Jungman:1994jr}. 
\end{itemize}

We show in the same figure an estimate of
the sensitivity of the neutrino telescope ANTARES (upper 
solid line). To estimate this sensitivity, we first 
evaluated the rate of muons in ANTARES from the direction of 
the GC, which depends (see, for example, Eq. (2.1) of
Ref.~\cite{Giesel:2003hj}) 
on specific experimental quantities, such as the detector 
effective area and the threshold energy for the detection of 
muons. The rate is higher for more energetic neutrinos, being 
proportional to the muon range and to the neutrino-nucleon cross 
section, which are both increasing functions of energy.

We found that ANTARES is most sensitive to the flux of neutrinos 
directly produced from $\bone$ annihilations. In fact, although 
the branching ratio for this channel is one order of magnitude 
smaller than the branching ratio into quark pairs, they are 
emitted at the highest available energy, $E=M$. To compute the 
sensitivity curve in our plot, we finally compared the predicted 
muon rate with the expected rate due to the atmospheric neutrino 
background, and thus estimated the required normalization for
our signal to exceed the background, for different values of the 
particle mass.

In Fig.~\ref{intflux} we also show the integral flux of photons (same 
threshold as for neutrinos, to compare relative flux) and the  
expected sensitivity of future experiments GLAST, MAGIC and 
HESS. 

We also note that, following \cite{Hooper:2002gs}, the flux 
of high energy neutrinos from $\bone$ annihilations in the sun 
should imply a muon rate around two orders of magnitude
larger than in the present case. Note also that our estimate 
for the $\gamma-$ray flux is in good agreement with Fig.~3 
of~\cite{Cheng:2002ej}. 

\section{Synchrotron Radiation}
\label{sy}

Another interesting mean of indirect detection of dark matter
is the synchrotron radiation originated from
the propagation of secondary $e^{\pm}$ in the Galactic magnetic
field. 

The magnetic field is supposed at equipartition (for details 
see \cite{melia,Bertone:2001jv}) in the inner part of the 
Galaxy and constant elsewhere. More specifically
\begin{equation}
B(r) = \mbox{max}\left[ 324 \mu \mbox{G} \left(\frac{r}{\mbox{pc}}\right)^
{-5/4}, 6\mu\mbox{G}\right]
\end{equation}
which means that the magnetic field is assumed to be in equipartition
with the plasma out to a galactocentric distance $r_c=0.23\,$pc,
and to be equal to a typical value observed throughout the Galaxy
at larger distances.

If the actual value of the magnetic field away from the central region
was smaller than the one we considered, this would imply a shift of
the radio spectrum
towards lower energies and thus, in the range of frequencies 
we are interested in,  a higher flux for a given frequency. 
This would also translate into stronger constraints for the mass and 
annihilation cross section. Nevertheless we prefer to be conservative
and consider a quite high value of $B$. Note that magnetic fields
stronger than equipartition values are physically unlikely.

The synchrotron flux per solid angle at a given frequency $\nu$
(cfr. Eq.~(22) in \cite{Bertone:2001jv}) is 
\begin{equation}
L_{\nu}(\psi) \simeq \frac{1}{4 \pi}\frac{9}{8}
\left(\frac{1}{0.29 \pi} \frac{ m_e^3 c^5}{e} 
\right)^{1/2} \frac{\sigma v}{M^2} Y_e(M,\nu)  \;\nu^{-1/2} \; I(\psi) 
\,,\label{synchrolum}
\end{equation} 
where 
\begin{equation} 
I(\psi) = \int_0^{\infty} ds \;\; \rho^2\left(r(s,\psi)\right) B^{-1/2}
\left(r(s,\psi)\right)\,,
\label{syn}
\end{equation} 
and $s$ is the coordinate running along the line of sight. $Y_e(M,\nu)$ is
the average number of secondary electrons above the energy $E_m(\nu)$ of
the electrons giving the maximum contribution at a given frequency
$\nu$ and for a magnetic field $B$. We recall that (Eqs.~(6) and (7) in 
\cite{Bertone:2001jv})
\begin{equation}
E_m(\nu) = \left( \frac{4 \pi}{3} \frac{m_e^3 c^5}{e} \frac{\nu}{B}
\right)^{1/2}\,.
\label{em}
\end{equation}
For frequencies around $400\,$MHz, used below, and for the lowest value
of the magnetic field, we find that $E_m(400\,\mbox{MHz})
\lesssim 2\,\mbox{GeV}$.
In reality, for dark matter profiles with central cusps, e.g. the
NFW, Kravtsov, and Moore profiles discussed above, most of the annihilation 
signal comes from the inner region of the Galaxy, where the magnetic field 
is probably higher. For $\nu=400\,\mbox{MHz}$ and $r < r_c$,
\begin{equation}
E_m(\nu)\simeq 0.3 \left(\frac{\nu}{400\,{\rm MHz}}\right)^{1/2}
\left(\frac{r}{\mbox{pc}}\right)^{5/8} \mbox{GeV}\,,\label{ecritic}
\end{equation}
which at the inner edge of the profile, corresponding to the Schwarzschild 
radius of the SMBH at the GC, $R_S=1.3\times10^{-6}$ pc, takes the value
$E_m(400\,\mbox{MHz})= 2.2 \times 10^{-5}$ GeV. We thus always have 
$E_m(400\,\mbox{MHz}) \ll M$, which means that most of the secondary
electrons are produced above this energy and contribute to the radio flux.

For a particle of mass $M$, the average electron multiplicity
per annihilation $Y_e(M)$ is
evaluated by adding the contribution of every annihilation channel $i$,
with cross section $(\sigma v)_i$, producing $Y_e^i(M)$ electrons
\be
\sigma v Y_e(M)=\sum_i (\sigma v)_i Y_e^i(M)\,,
\label{sigY}
\ee
where $\sigma v$ is again the total annihilation cross section.

The main channels contributing to this flux are direct production of 
leptons, and annihilation into quarks, as discussed above. In the 
first case, we have $Y_e^{e^{\pm}}(M)=Y_e^{\mu^{\pm}}(M) \simeq 2$ in all
the relevant range of masses, while in the quark channel, to count the number 
of electrons $Y_e^{q\overline{q}}(M)$, we integrate the FF 
for $e^+$ shown in Fig.~(\ref{gaenu}), and multiply by a factor 2 to
account for $e^-$ from the corresponding decay of $\pi^-$. Using
Eq.~(\ref{sigY}) and
extrapolating the FF in Fig.~(\ref{gaenu}) at low $x$ values, we find e.g.
$\sigma v Y_e(1\,\mbox{TeV})\simeq 6 \times 10^{-3}\,\mbox{TeV}^{-2}$,
and $Y_e(1\,\mbox{TeV})\simeq 4.5$ for $M=1\,$TeV. The electron
multiplicity in the hadronic channel alone would be much larger,
namely roughly 20.

To obtain the observed radiation, one should multiply the luminosity 
$L_\nu$ with the synchrotron self-absorption coefficient. 
In our case optical depth is negligible and the self-absorption
coefficient of the order of unity. In fact, using the expression
introduced in \cite{Bertone:2001jv}, the optical depth can be expressed as
\begin{equation}
\tau\simeq\frac{\sigma v}{M^2} \frac{Y_e(M)}{4 \pi} \frac{1}{\nu^3} 
\int_0^{d_\odot} ds\rho^2(s)\,,
\end{equation}
where $d_\odot\simeq8\,$kpc is the distance of the sun from the GC.
Using $M=1\,$TeV, $\sigma v = 1.6\times10^{-4}\,$TeV$^{-2}$ (cross
section for annihilation into right-handed up quarks) and a
NFW profile we find 
$ \tau = 1.78 \times 10^{-4} \left( \nu / 100\mbox{MHz} 
\right)^{-3}$. We can thus neglect self-absorption unless
the frequency considered is very small. The absorption
on relativistic electrons from other sources is also 
negligible: Using $n(E) \lesssim 10^{-2}\,{\rm GeV}^{-1}{\rm cm}^{-2}
{\rm s}^{-1}{\rm sr}^{-1}$ for the locally observed differential electron
flux in the relevant energy range given by Eq.~(\ref{ecritic})~\cite{ms},
one obtains for the absorption coefficient per length
$\alpha_\nu\lesssim 6 \times 10^{-16} \mbox{pc}^{-1}
(B/\mu\mbox{G})(\nu/\mbox{GHz})^{-2}$. Obviously, even if the
relativistic electron flux due to non-acceleration processes close
to the GC is orders of magnitude larger, the effect would still be
negligible. However, for frequencies below
a few MHz free-free absorption sets in (see e.g. \cite{cane}).

\begin{figure}
\includegraphics[width=0.48\textwidth,clip=true]{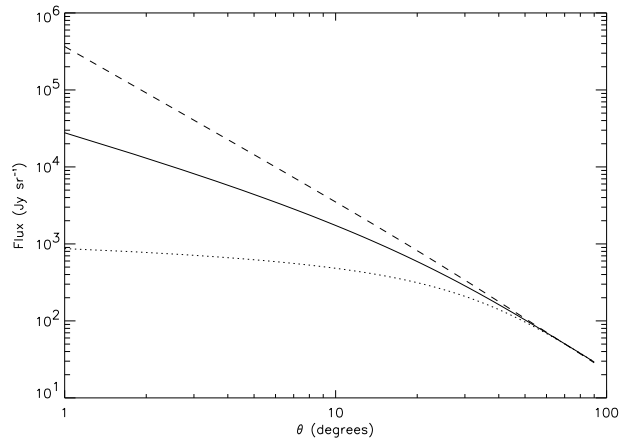}
\caption{Synchrotron emission at frequency $\nu=327\,$MHz for
NFW (solid line), Moore (dashed) and Kravtsov (dotted) profiles,
as a function of the angular distance from the Galactic center.}
\label{ang}
\end{figure}

In Fig.~\ref{ang} we evaluate Eq.~(\ref{synchrolum}) for three different
halo profiles, as a function of the angular distance from the GC. 

To compare with observations we integrate over the relevant solid angle.
The comparison between predicted and observed fluxes constrains the 
cross section of annihilating dark matter particles and therefore their 
mass. We studied three different cases:

\begin{enumerate}
\item Flux at $\nu=408\,$MHz in a cone of half-width 4 arcsec
pointing towards the GC. Assuming a NFW profile, the 
comparison with the observed flux, which is $\lesssim 0.05\,$Jy~\cite{davies},
puts the following constraint on the cross section
\be
\sigma v\lesssim1.5\times 10^{-24}\left(\frac{M}{\mbox{\small{TeV}}} \right)^2 
\frac{Y_e^{kk}(1\mbox{\small{TeV})}}{Y_e(M)}\, \mbox{cm}^{3} \mbox{s}^{-1}
\ee
where $Y_e^{kk}$ refers to the $\bone$ particle, while
$Y_e$ represents the electron yield for a generic candidate.

\item Flux at $\nu=327\,$MHz from a circular ring around the GC
 with inner and outer radius equal to 5 and 10 arcmin, respectively.
We estimated the observed flux from~\cite{ananta} to be
$\simeq121\,$Jy. If we impose the predicted
flux for a NFW profile not to exceed the observations, then
\be
\sigma v\lesssim6.0\times 10^{-24}\left(\frac{M}{\mbox{\small{TeV}}}\right)^2 
\frac{Y_e^{kk}(1\mbox{\small{TeV})}}{Y_e(M)}\,\mbox{cm}^{3} \mbox{s}^{-1}. 
\ee

\item Flux at $\nu=327\,$MHz in a cone of half-width 13.5 arcmin
pointing towards the GC. The observed flux is $\simeq 362\,$Jy~\cite{lazio}.
The constraint on the cross section in this case is (for a NFW profile) 
\be
\sigma v\lesssim6.9\times 10^{-24}\left(\frac{M}{\mbox{\small{TeV}}}\right)^2 
\frac{Y_e^{kk}(1\mbox{\small{TeV})}}{Y_e(M)}\,\mbox{cm}^{3} \mbox{s}^{-1}. 
\ee

\end{enumerate}

\begin{figure}
\includegraphics[width=0.48\textwidth,clip=true]{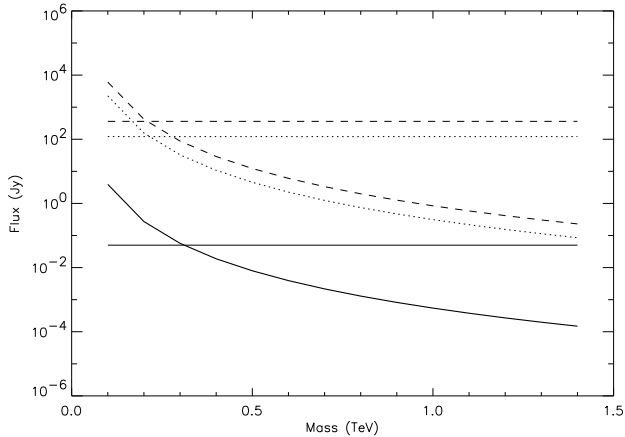}
\caption{Predicted (curves) and observed (horizontal lines) radio flux 
from regions close to the Galactic center, for a NFW profile, as a 
function of the particle 
mass. Bottom to top: case 1 (solid lines), case 2 (dotted lines) and
case 3 (dashed lines), see text for details.}
\label{synchro}
\end{figure}

The constraints derived on the cross section are quite general and
apply to any type of self-annihilating dark matter particles. 
To test specific candidates one has to specify
the relation between cross section and mass. In particular the above constraints
can be turned in a lower bound on the mass of the KK
particle. In Fig.~\ref{synchro} we show predicted and observed 
fluxes for KK particles, for a NFW profile, as a function of the particle 
mass. For every case we plot the predicted and observed flux, the latter being
of course represented by a horizontal line. Cases 1, 2 and 3 are
respectively represented by solid, dashed and dotted lines. 
Case 1 is the most constraining, implying a lower bound on the
mass of about $0.3\,$TeV. 

\begin{figure}
\includegraphics[width=0.48\textwidth,clip=true]{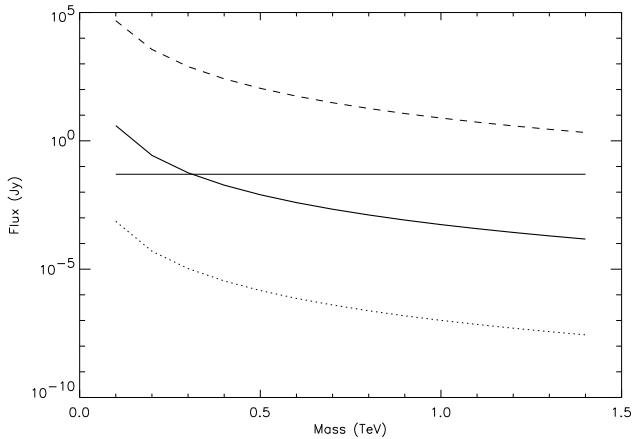}
\caption{Predicted radio flux from a 4 arcsec cone around the Galactic
center, as a function of the KK particle 
mass for different density profiles. Bottom to top: Kravtsov et al. 2nd example
(dotted line), NFW (solid line) and Moore et al (dashed line).}
\label{synchro2}
\end{figure}

To emphasize the importance of the density profile adopted, we plot
in Fig.~\ref{synchro2} the flux corresponding to case 1 for three
different profiles. It is evident that for a Moore et al. profile
the synchrotron flux would exceed the observed emission by several
orders of magnitude for any interesting value of the $\bone$ mass,
while for a Kravtsov profile $\bone$ particles are practically
unconstrained.

We also compared high latitude predicted fluxes with 
observations \cite{cane}. The strongest constraints result from
the lowest frequencies at which free-free and synchrotron self-absorption
are not yet important, i.e. around $10\,$MHz~\cite{cane}. Here,
the observed background
emission between $0^\circ$ and $90^\circ$ from the Galactic
anti-center is $\simeq6\times10^6\,$Jy. Comparing with the
predicted emission results in the limit
\be
\sigma v \lesssim10^{-22}\left(\frac{M}{\mbox{\small{TeV}}} \right)^2 
\frac{Y_e(1\mbox{\small{TeV})}}{Y_e(M)}\,\mbox{cm}^{3} \mbox{s}^{-1}. 
\ee
While this is considerably weaker than the constraints above,
it is largely independent of the unknown GC dark
matter profile.

Note that we can safely neglect other energy losses of secondary electrons
such as Inverse Compton Scattering (ICS) and Pair Production (PP), which are
much less efficient than synchrotron emission at the GC. Consider
in fact the Synchrotron energy loss time 
\be
t_e\simeq E \left( \frac{dE}{dt} \right)^{-1} =
\frac{3m_e^4}{2 e^4 B^2 E}\,.
\ee
This implies a synchrotron photon energy density
$u \simeq t_e \sigma v \rho^2/M \lesssim 10^{-51}$GeV$^4$,
compared to the cosmic microwave background energy density
$u_{CMB} \simeq 10^{-51}$GeV$^4$. If $\sigma_T$ is the Thomson
cross section, we can express the ICS loss time as (see e.g. \cite{rybi}) 
\be
t_{ICS} \simeq \frac{3 m_e^2}{4 \sigma_T E u} \gtrsim 8 \times 10^{12} 
\left( \frac{\nu}{\mbox{MHz}}\right)^{-1/2} \mbox{pc}\,.
\ee
The energy loss time is even longer for PP, $t_{e^+e^-} \gtrsim 8.8 \times
10^{29} \mbox{pc}$.

Finally we can check whether we can get interesting constraints 
from clumped halos. In fact, high-resolution N-body simulations suggest 
the existence of many sub-structures in the dark halos (see e.g.
\cite{Moore:1999gc}).
Without going into details, we refer to \cite{Blasi:2002ct}, where  
synchrotron emission is evaluated for neutralino annihilation 
in clumped halos. 

If we consider clumps with a NFW profile, we can refer to Fig.~2
of~\cite{Blasi:2002ct} and extrapolate the flux for a 1TeV neutralino
at $\nu=0.1\,$GHz to be $\lesssim10\,$Jy, in a solid angle which we
estimate, using the normalization of
CMB anisotropies, to be around $\Delta\Omega=3\times10^{-3}\,$sr.
We can have a rough idea of the flux corresponding to our clump
candidate, for the same mass,
by rescaling the flux by the ratio of the $\bone$ to the neutralino
annihilation cross section. The expected flux for our candidate clump
should thus be of the order of $100\,$Jy. 

To see if this flux can outshine the background, we compare it
with low frequency observations of the galactic anti-center regions
for which recent estimates suggest a diffuse flux, at $\nu=0.1$ GHz, of
$3 \times 10^5\,$Jy sr$^{-1}$~\cite{radiobook}, which for the above
solid angle would give a flux of $\approx 10^3\,$Jy, 
thus exceeding the predicted flux by one order of magnitude.
Thus the hypothesis of a clumped halo for our Galaxy does not
put any further constraint on
the parameters of our dark matter candidate. 

The above constraints  would become much more 
stringent if a spike extended at the GC, nevertheless in 
this case self-absorption would play a more important role
, requiring a careful analysis~\cite{Bertone:2001jv}.
 
\section{Conclusions}
\label{conc}
We have evaluated the prospects for indirect detection of $\bone$,
the first Kaluza-Klein state of the Hypercharge $B$ gauge boson. In particular,
we focused on neutrino, $\gamma-$ray, and synchrotron fluxes from
annihilation of $\bone$ particles in the Galactic halo.

Assuming a Galactic magnetic field at equipartition in the inner
part of the Galaxy and equal a few micro Gauss elsewhere, we showed that,
with the same density profile, synchrotron radiation
give stronger constraints than $\gamma$-ray emission,
for what concerns existing observations,
while neutrino fluxes are much below the expected sensitivities 
of future experiments and several orders of magnitude smaller with 
respect to those expected from annihilations in the Sun.

Significant constraints on $\gamma-$rays will result from
hypothetic null searches at the expected sensitivity of future
experiments like GLAST, MAGIC and HESS. In particular, a null 
search of MAGIC would imply a lower bound on the mass of
$M\gtrsim0.6\,$TeV for a NFW profile. The existence of a spike at the
GC would enhance the observed fluxes by several orders of 
magnitude, and would thus basically rule out the whole
range of masses.

On the other hand, predicted  synchrotron fluxes are less 
robust because they not only depend on the dark
matter profile close to the Galactic center, but also on specific assumptions
for the Galactic magnetic field. 

Comparing predicted and observed synchrotron fluxes from
the Galactic center, we derived, for a NFW profile, 
an upper bound for the annihilation cross section of
$\sigma v \le 1.5\times 10^{-24}(M/\mbox{TeV})^2
Y_e^{kk}(1\,\mbox{TeV})/Y_e(M)\,$cm$^3$ s$^{-1}$, which translates, for 
$\bone$ particles, into a lower bound on the mass of about 
$M\gtrsim0.3\,$TeV. We also discussed how this bound would vary 
depending on the dark matter profile adopted. 

To conclude, the Minimal Supersymmetric Standard Model is not the only
viable extension of the Standard Model and the LSP is not 
the only viable Dark Matter candidate. It is therefore important to
be open to other alternatives. Here, we focused on the Lightest
Kaluza--Klein Particle, $\bone$, a typical WIMP and a viable DM
candidate. Direct and indirect detection
prospects for $\bone$ are challenging. In this work, we were able to set a 
constraint on the mass of the LKP and therefore on the compactification scale 
of Universal Extra Dimensions. Such a constraint comes from an existing
measurement of synchrotron radiation and is of the same order as the
one from current electroweak precision tests. However, a better
knowledge of the Galactic Dark Matter profile together with that of
the magnetic field would be helpful to put this constraint on a more
robust footing. The idea of TeV Kaluza--Klein Dark Matter remains
safe for a while.

\section*{Acknowledgements}
This work is supported in part by the US Department of 
Energy, High Energy Physics Division, under contract 
W-31-109-Eng-38, by the David and Lucile Packard Foundation and
also by the EU Fifth Framework Network "Supersymmetry and the 
Early Universe" (HPRN-CT-2000-00152).
GB and GS would like to thank Joe Silk for many valuable discussions.
GS thanks Carlos de Breuck for illuminating informations on radio
observations.

\end{document}